\documentclass[10pt,twocolumn,nofootinbib,showpacs]{revtex4}
\usepackage{amssymb,amsbsy,amsmath,amsfonts,amssymb,amscd}
\usepackage{latexsym,euscript,exscale,epic,eepic,epsfig}
\usepackage[utf8]{inputenc}
\usepackage[english]{babel}
\usepackage[unicode=true,
 bookmarks=true,bookmarksnumbered=false,bookmarksopen=false,
 breaklinks=true,pdfborder={0 0 0},backref=false,colorlinks=true,citecolor=blue,linkcolor=red]{hyperref}
\hypersetup{pdftitle={Isotropization of the universe during inflation},
 pdfauthor={Thiago Pereira, Cyril Pitrou}}

\usepackage{times}
\usepackage{lipsum}
\usepackage{color}

\newcommand*{\mycdot}{\raisebox{-0.25ex}{\scalebox{1.2}{$\cdot$}}}
\newcommand{\be}{\begin{equation}}
\newcommand{\ee}{\end{equation}}
\def\stau{{\hat{\sigma}}}

\def\dd{{\rm d}}

\begin{document}

\title{Isotropization of the universe during inflation}
\author{Thiago S. Pereira}
\email{tspereira@uel.br}
\affiliation{  Universidade Estadual de Londrina (UEL)\\ 
  Rodovia Celso Garcia Cid, km 380, 86057-970, Londrina -- PR, Brasil}
\author{Cyril Pitrou}
\email{pitrou@iap.fr}
\affiliation{CNRS, Institut d’Astrophysique de Paris (IAP), \\
Sorbonne Universités, UPMC Univ Paris 6,\\ 98bis, boulevard Arago, 75014 Paris, France}


\begin{abstract}
{A primordial inflationary phase allows one to erase any possible anisotropic expansion thanks to 
the cosmic no-hair theorem. If there is no global anisotropic stress, then the anisotropic expansion 
rate tends to decrease. What are the observational consequences of a possible early anisotropic 
phase? We first review the dynamics of anisotropic universes and report analytic approximations. We 
then discuss the structure of dynamical equations for perturbations and the statistical properties 
of observables, as well as the implication of a primordial anisotropy on the quantization of these 
perturbations during inflation. Finally we briefly review models based on primordial vector field 
which evade the cosmic no-hair theorem.}
\end{abstract}

\maketitle


\section{Introduction}
For good or for evil, inflation is currently the only known mechanism capable of explaining the 
origin and statistical properties of large scale structures in the universe. Given its central 
role on the standard cosmological model, it is thus important to test its robustness 
in all possible ways. 

Even though inflation was designed, among other things, to wash away classical 
inhomogeneities~\cite{Guth:1980zm}, most of its implementations start with a symmetric background 
from the onset. This approach was mainly supported by large field models in which a long 
period of exponential expansion takes place~\cite{Linde:2007fr}, thus effectively erasing all 
possible memories of initial conditions. However, if inflationary models predicting a small number 
of e-folds turn out to be favoured, then the state of the universe before the onset of inflation  
can play an important role for cosmological observables, and thus they need to be included in a 
self-consistent way.

From an observational perspective, the question of the relevance of pre-inflationary initial 
conditions was boosted by the detection of large-scale statistical anomalies in the cosmic 
microwave background (CMB) temperature maps~\cite{Bennett:2010jb,Ade:2013nlj,Ade:2015hxq}. If 
inflation is sensitive to the initial conditions such as spatial inhomogeneities, 
then these features could be imprinted on the primordial power spectrum at large scales, and thus 
possibly related to the origin of CMB anomalies. 

As it turns out, the implementation of inflation in a broader geometrical framework leads to 
several important questions, including the very possibility of an inflating universe in the presence 
of large inhomogeneities~\cite{Goldwirth:1990pm,Deruelle:1994pa}. In this review paper, we address 
the implementation of slow-roll inflation without assuming some of the symmetries that the 
mechanism is supposed to predict. Specifically, we investigate the dynamics of the inflaton 
when released in an spatially homogeneous {\it but anisotropic} spacetime of the Bianchi I family. We 
show that the existence of an adiabatic (Bunch Davies) vacuum cannot be guaranteed throughout the 
anisotropic phase, and thus that the amplitude of strongly anisotropic modes cannot be unambiguously 
fixed. 

This review is organized as follows: we start by recalling the basic equations and definitions of
Bianchi I spacetimes in \ref{SecBackGen} We then use these equations to find analytical solutions for the 
geometry in the presence of a cosmological constant in \ref{desitter}. Using these solutions, we 
conduct a semi-analytical investigation of the inflationary dynamics using the chaotic potential as a proxy 
in \ref{chaoticmodel}. In \ref{quantprocedgen} we discuss the quantizations procedures 
(\ref{quantproced1}), the general properties of linear perturbations (\ref{perturb}), and the 
connection between anisotropic inflation and CMB anomalies (\ref{cmbanomalies}). In \ref{alternative} 4 
we briefly comment on some alternative models, including vector and shear-free anisotropic 
inflation. 

Throughout this text, Latin/Greek indices refer to space/spacetime coordinates. Moreover, we adopt 
the convention in which indices separated by square brackets are not summed over. So, for example, 
$[a_i]\delta^i_j$ carries no sum in $i$, whereas $a_i\delta^i_j$ does.

\section{Inflation in Bianchi I spacetimes}

\subsection{Background generalities}\label{SecBackGen}

We start by studying inflation in Bianchi I (BI) spacetimes. These are exact solutions of 
Einstein equations describing homogeneous but anisotropically expanding spacetimes with flat 
spatial sections. In comoving coordinates, their metric is given by
\begin{align}
\label{background-g}
{\dd}s^2 = -\dd \tau^2 + S^2(\tau)\gamma_{ij}(\tau)\dd x^i\dd x^j\,, 
\end{align}
where $\tau$ is the time measured by comoving observers, $\gamma_{ij}$ is the metric on 
constant-time hypersurfaces, parametrized by
\be
\gamma_{ij}(\tau) = \exp[2\beta_i(\tau)]\delta_{ij}\,.
\ee
$S$ is the (geometrical) average of the three individual scale factors $X_i \equiv Se^{\beta_i}$, 
since
\be
\label{Smean}
S\equiv(X_1X_2X_3)^{1/3}.
\ee
The functions  $\beta_{i}(\tau)$ are not independent, but constrained by 
\be\label{trace-free}
\sum_{i=1}^3\beta_i = 0\,,
\ee
which ensures that spatial comoving volumes are constant in time ($\sqrt{\gamma}=1$). 
Anisotropic expansion induces a geometrical shear, which is described by the \emph{shear 
tensor} and \emph{shear scalar}
\be
\label{shear-def}
\stau^i_j \equiv [\dot{\beta}_i]\delta^i_j\,,\quad \stau^2\equiv\stau^i_j\stau^j_i\,,
\ee
where a dot means $\dd/\dd\tau$ and spatial indices are lowered by
$\gamma_{ij}$ and raised with its inverse $\gamma^{ij}$.

In the presence of a perfect fluid of energy density $\rho$ and pressure $p$, the background 
Einstein equations are\footnote{ $\kappa\equiv8\pi G\equiv M_p^{-2}$.}
\begin{align}
3H^2 & = \kappa\rho +\frac{1}{2}\hat{\sigma}^2\,,\label{H2} \\
\frac{\ddot S}{S} & = -\frac{\kappa}{6}(\rho+3p)-\frac{1}{3}\hat{\sigma}^2\,, \\
(\hat \sigma^i_j)^{\mycdot} & = -3H\hat{\sigma}^i_j\,.\label{sheareq}
\end{align}
where $H \equiv\dot{S}/S$ is the (average) Hubble parameter. These
equations can be combined to recover the fluid conservation equation
$\dot{\rho}+3H(\rho+p)=0$. Since there are no sources for the stress dynamics, Eq.~\eqref{sheareq} implies 
that the shear has only a decaying mode
\be
\label{shear-decay}
\stau^i_j = \frac{[C_i]}{S^3}\delta^i_j\,,\quad\stau^2 = \frac{C^2}{S^6}\,.
\ee
The constants $C_i$ can be parametrized by
\be
C_i = \sqrt{\frac{2}{3}}C\sin\alpha_i\,,\quad\alpha_i = \alpha+i \frac{2\pi }{3}\,,
\ee
which automatically ensures that $\sum_iC_i=0$ and $C^2=\sum_iC_i^2$. Clearly, $C$ measures the 
initial amplitude of the shear. The constant $\alpha$, on the other hand, represents a residual 
freedom in the choice of the initially expanding/contracting eigendirections of the shear. 
Before moving on, we note from Eqs.~\eqref{shear-def} and~\eqref{shear-decay} that $\beta_i$ can be 
written as
\be
\label{betai}
\beta_{i}(\tau) = C_i\int^\tau \frac{\dd\tau'}{S^3(\tau')}\,.
\ee
This expression will be useful in finding analytical solutions of the
anisotropic phase.

\subsection{``de Sitter'' expansion}\label{desitter}
\begin{figure}
\begin{centering}
\includegraphics[scale=0.45]{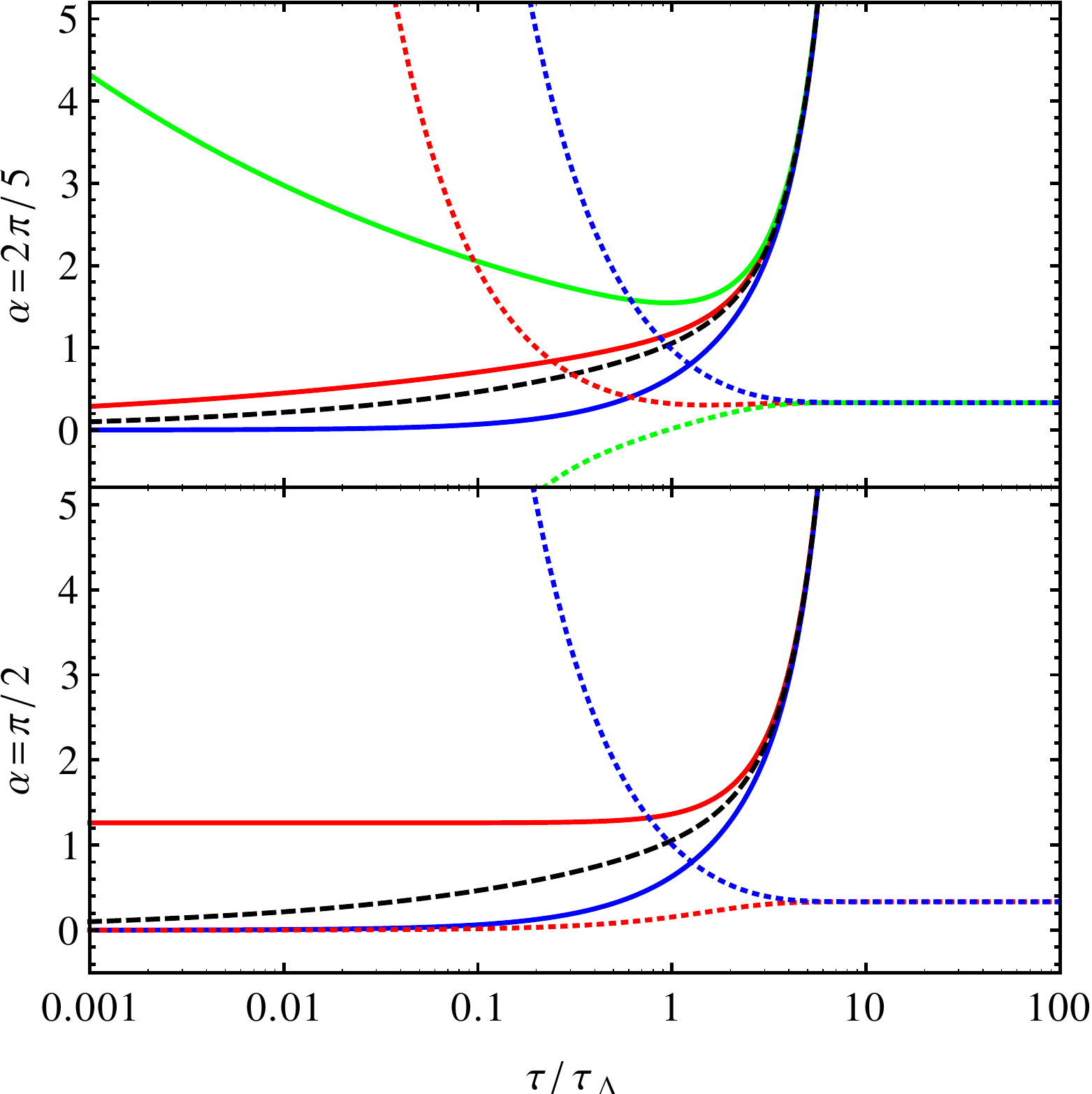}
\par\end{centering}
\protect\caption{Three directional scale factors (thick, continuous lines) as a function of time 
for two values of $\alpha$. Note that they all converge to the mean scale factor (thick, dashed 
line). We also show the directional Hubble parameters, $\dot{X}_i/X_i$ (thin, dotted 
lines).}\label{scalefacfig}
\end{figure}

Before going into the details of slow-roll inflation in BI spacetimes, it is rewarding to analyse 
the expansion in the presence of a pure cosmological constant, for which solutions can be found 
analytically~\cite{Pitrou:2008gk}. These solutions keep most of the
main features that are found in the more general (slow-roll inflation) case. From the behaviour of Eqs.~\eqref{H2} 
and~\eqref{shear-decay}, we see that the universe starts from a shear-dominated phase, followed by a 
(nearly) de Sitter expansion. Setting $\kappa\rho=\Lambda$ in Eq.~\eqref{H2} and integrating, we 
find that~\cite{Moss:1986ud,Pitrou:2008gk}
\be
\label{slambda}
S(\tau) =\left[\sqrt{\frac{3}{2}}C\tau_\Lambda\sinh(\tau/\tau_\Lambda)\right]^{1/3}\,, 
\ee
with the typical time scale $\tau_\Lambda\equiv(3\Lambda)^{-1/2}$. The average Hubble parameter then becomes
\be
\label{hlambda}
H=\frac{1}{3\tau_\Lambda}\coth\left(\frac{\tau}{\tau_\Lambda}\right)\,.
\ee
For $\tau\gg\tau_\Lambda$, $\coth(\tau/\tau_\Lambda)$ approaches 1 and we recover 
$3H^2\approx\Lambda$, as expected. 
Thus, $\tau_\Lambda$ is the typical duration of a primordial anisotropic phase prior to a ``pure'' 
de Sitter expansion. The isotropization is better illustrated by plotting the directional scale 
factors, which can be obtained by integrating Eq.~\eqref{betai}
\be
\beta_i(\tau) =  
\log\left[\tanh\left(\frac{\tau}{2\tau_\Lambda}\right)\right]^{\frac{2}{3}\sin\alpha_i}
\ee
and using the definition~\eqref{Smean}. With this, we find  that the individual scale factors
evolve as
\be
\label{directionalX}
X_i=S(\tau)\left[\tanh\left(\frac{\tau}{2\tau_\Lambda}\right)\right]^{\frac{2}{3}\sin\alpha_i}\,.
\ee
We plot in Figure~\ref{scalefacfig} the typical behaviour of these scale factors, on which the 
isotropization process is obvious. Because of the condition~\eqref{trace-free}, there will always 
exist one bouncing direction, except for the case of $\alpha=\pi/2$. The case $\alpha=\pi/2$ is 
necessarily exceptional since it is the only model for which the invariant 
$R_{\alpha\beta\mu\nu}R^{\alpha\beta\mu\nu}$ is finite as $\tau\rightarrow0$. Since this invariant 
diverges at the singularity for any other $\alpha$, we conclude that the case $\alpha=\pi/2$ is 
singular~\cite{Pitrou:2008gk}. In fact, in absence of a cosmological constant ($\Lambda=0$), it also 
corresponds to a patch of the Minkowski spacetime~\cite{Kofman:2011tr}.


\subsection{Slow-roll inflation}\label{chaoticmodel}

We are primarily interested in the inflationary dynamics on a BI spacetime geometry. We thus assume 
that the inflaton is described by a single, canonical scalar field with energy-momentum tensor 
\[
T_{\mu\nu} = 
\partial_\mu\varphi\partial_\nu\varphi-\frac{1}{2}
\left(\partial_\lambda\varphi\partial^\lambda\varphi+2V\right)g_{\mu\nu}\,.
\]
For simplicity we will focus on the model of chaotic inflation~\cite{Linde:1983gd}
\be
V(\varphi) = \frac{1}{2}m^2\varphi^2\,.
\ee
We stress however that most of our results can be easily extended to other potentials. 

For a homogeneous field, $\varphi=\varphi(\tau)$, the Klein-Gordon equation involves only 
the trace of the spatial metric. Thus, the dynamics of $\varphi$ is formally the same as in 
Friedmann-Lema\^itre (FL), i.e. isotropic universes, that is
\be
\label{kgeq}
\ddot{\varphi}+3H\dot{\varphi}+V_{,\varphi}=0\,.
\ee
However, we must stress that the (average) Hubble parameter is affected by the presence of shear, 
which indirectly affects $\varphi$. In what follows it will be convenient to define a dimensionless 
parameter $x\equiv \stau/\sqrt{6}H$. In terms of $x$, Eq.~\eqref{H2} becomes
\be
(1-x^2)H^2 = \frac{\kappa}{3}\rho
\ee
and it implies that $x<1$ in order to ensure that the energy density is positive. We also introduce 
two slow-roll parameters as follows
\be
\epsilon\equiv \frac{3}{2}\dot{\varphi}^2\left[\frac{1}{2}\dot{\varphi}^2+V\right]^{-1},\label{sr1}
\quad\delta\equiv -\frac{\ddot\varphi}{H\dot\varphi}\,.
\ee

\begin{figure}[ht]
\begin{centering}
\includegraphics[scale=0.54]{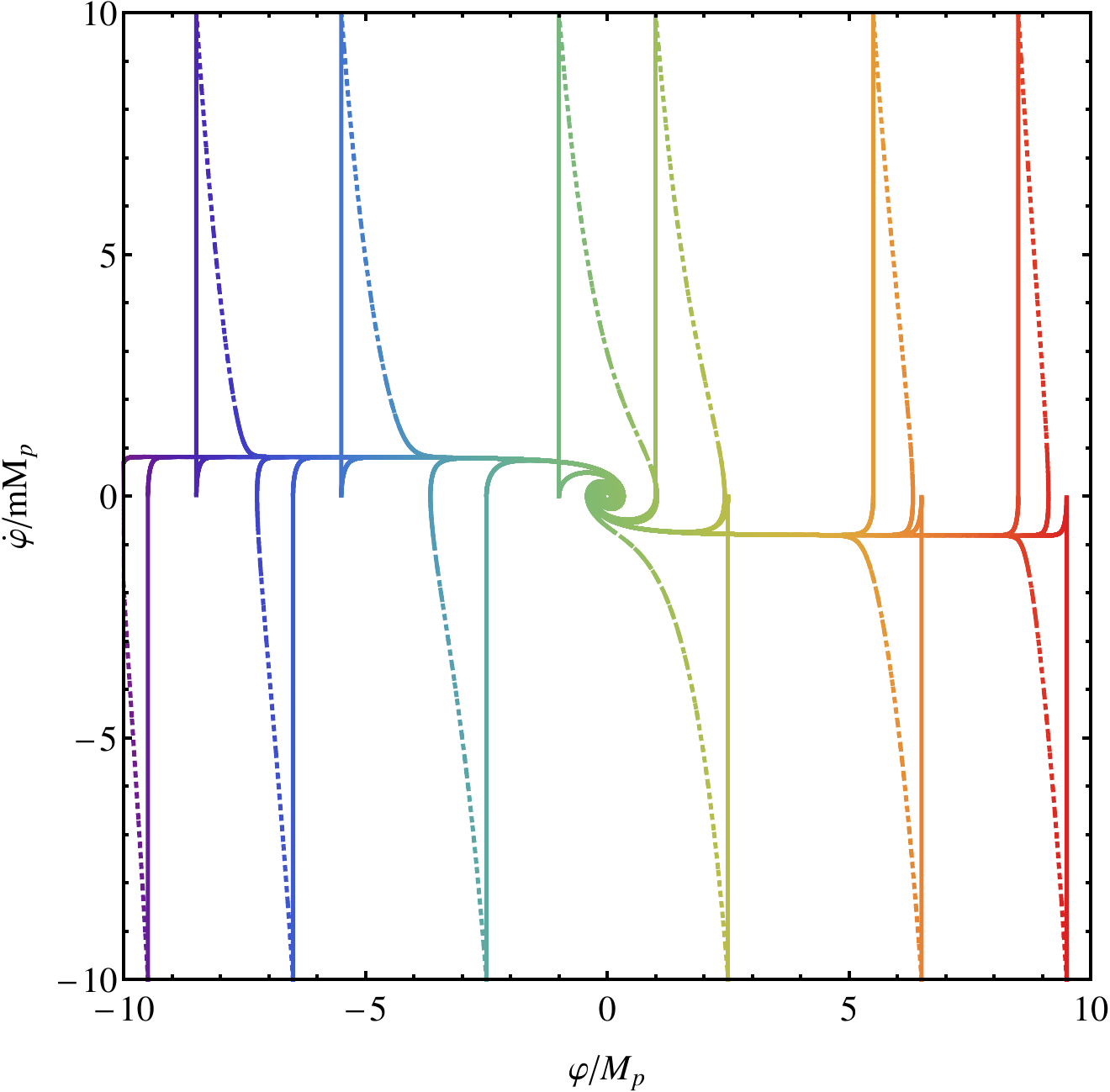}
\par\end{centering}
\protect\caption{Attractor behaviour of the inflaton in an anisotropic universe (continuous lines). 
During shear domination, $\dot\varphi$ decays steadily, while $\varphi$ remains constant. This 
behaviour is followed by the slow-roll regime once the shear is sufficiently small. For comparison 
we also show the inflaton behaviour in a FL background with the same initial conditions for 
$\varphi$ and $\dot{\varphi}$ (dotted lines).}\label{phasespace}
\end{figure}

In standard inflation, the dynamics of the universe is characterized by an attractor 
(slow-roll) regime in which both $\epsilon$ and $\delta$ are small and their time derivatives 
behave as ${\cal O}(\epsilon^2,\epsilon\delta)$~\cite{peter2013primordial}. The main effect of the 
spatial anisotropy is to introduce a second attractor regime in the inflationary dynamics. Indeed, 
close to the singularity the shear dominates, and the solutions~\eqref{slambda}-\eqref{hlambda} are 
a good description of the evolution of the universe. Assuming that $V\gg\dot{\varphi}^2$, we 
find\footnote{If $\dot{\varphi}^2\gg V$, then $\dot{\varphi}\sim S^{-3}$ from Eq.~\eqref{kgeq}. At 
early times it decreases as $\dot{\varphi}\sim\varphi_0/\tau$, so that it quickly converges to the 
slow-roll regime.} ~\cite{Gumrukcuoglu:2007bx,Pitrou:2008gk}
\begin{align*}
\varphi(\tau) & = 
\varphi_0\left[1-\frac{1}{6}\left(\frac{M_p}{\varphi_0}\right)^2\frac{\tau^2}{\tau_0^2}
+{\cal O}\left(\frac{\tau^4}{\tau_0^4}\right)\right]\,, \\ 
\delta(\tau) & = 
-3\left[1-\frac{1}{2}\frac{\tau^2}{\tau_0^2}+{\cal 
O}\left(\frac{\tau^4}{\tau_0^4}\right)\right]\,, \\
\epsilon(\tau) & = \frac{1}{3}\left(\frac{M_p}{\varphi_0}\right)^2\frac{\tau^2}{\tau_0^2}+{\cal 
O}\left(\frac{\tau^4}{\tau_0^4}\right)\,,
\end{align*}
where $\tau_0\equiv\sqrt{2/3}M_p/m\varphi_0$. 
We thus see that, during the shear-dominated regime, 
the solutions are attracted to the point $(\dot{\varphi},\varphi)=(0,\varphi_0)$ regardless of the 
initial conditions. Moreover, note that $\epsilon\rightarrow0$, but
$\delta\rightarrow-3$ 
during this regime. Actually, one can also show that 
$\dot\epsilon=2H\epsilon(\epsilon-\delta)\approx 6H\epsilon$, and  thus, contrarily to standard 
inflation, the time evolution of $\epsilon$ cannot be neglected when the shear 
dominates~\cite{Pitrou:2008gk}. However, once the shear becomes negligible we reach the standard 
slow-roll regime, and the universe inflates. This double attractor behaviour is illustrated in 
Figure~\ref{phasespace}.

Finally, one might worry that if the field moves by a large fraction during shear-domination, then 
the shear could affect the number of e-folds left during slow-roll inflation. However, as one can 
see from Figure~\ref{phasespace}, the fractional variation $(\varphi_0-\varphi_{\rm 
sl}[\varphi_0])/\varphi_0$, where $\varphi_{\rm sl}$ is the field value at the beginning of 
slow-roll expansion, decreases for increasing $\varphi_0$, so that the effect of the shear on the 
number of e-folds is negligible~\cite{Pitrou:2008gk}.

\section{Dynamics of fluctuations}\label{quantprocedgen}
\subsection{Quantization generalities}\label{quantproced1}

As we have seen, the effect of slow-roll inflation in a BI universe is to quickly erase classical 
anisotropies. This is indeed one of the primordial purposes of the inflationary mechanism. On the 
other hand, the existence of primordial anisotropic expansion drastically affects the evolution of 
quantum perturbations, so that in principle one expects to find signatures from the early 
anisotropic stage imprinted on the primordial power spectrum of CMB fluctuations.

Due to the lack of rotational invariance in BI spacetimes, the evolution of cosmological 
perturbations \cite{Pereira:2007yy} differs drastically from the one
found in FL universes \cite{Mukhanov:1990me}. The differences can be traced back to two main effects. First, 
the background shear tensor couples scalar, vector and tensor degrees of freedom already at the 
linear level of perturbations. This means that, apart from the scalar and tensor primordial power 
spectra, there are cross correlations between scalar and tensor modes and from tensor modes of 
different polarizations~\cite{Pereira:2007yy,Gumrukcuoglu:2010yc}. Second, owing to the spatial 
homogeneity of the BI manifold, any observable $f$ can be decomposed in terms of plane-waves
\be
f(x^i,\tau) = \int\frac{d^3k}{(2\pi)^{3/2}}\tilde{f}(k_i,\tau)e^{{\rm{i}}k_jx^j}\,.
\ee
However, since Fourier co-vectors $k_i$ are constant in time, their duals develop a time-dependence 
through the spatial metric as $k^i(\tau)=\gamma^{ij}(\tau)k_j$. In particular, two Fourier modes with the 
same co-moving norm $k$ can have quite different time evolutions depending on their initial 
directions, since now 
\be
\label{ksquare}
k^2=\sum_{i,j=1}^3(e^{-2\beta_i(\tau)}\delta^{ij})k_ik_j\,.
\ee
This implies that the power spectrum of a given observable will acquire a dependence on 
the direction of the vector $\mathbf{k}$. 

While the first of the two aforementioned effects lead to see-saw mechanisms which have important 
consequences for the formation of structures in the late 
universe~\cite{Pereira:2015jya,Pitrou:2015iya}, the second has immediate consequences for the 
quantization of inflationary perturbations. As shown in Eq.~\eqref{directionalX} (see also 
Figure~\ref{scalefacfig}), as we approach the early anisotropic stage there is always one spatial 
direction going through a bounce\footnote{Except for the singular case $\alpha=\pi/2$}. The growth 
of this direction as $\tau\rightarrow0$ implies that the wavelength of perturbations will 
eventually cross the (mean) Hubble horizon. 

Consider for example a {\emph{comoving} mode aligned with a (fixed) direction $i$. 
From Eq.~\eqref{ksquare} we have
\be
\frac{k}{SH}=\frac{k_i}{HX_i}\sim\tau^{2(1-\sin\alpha_i)/3}\,,
\ee
where we have used the solutions of Sec.~\ref{desitter}. Since the power in $\tau$ is positive for 
any $\alpha_i$, we conclude that any given mode will exceed the Hubble scale when the shear 
dominates. 
The amplitude of such a mode cannot be fixed unambiguously by standard quantization procedures, and 
inflation is expected to lose predictability during this phase. 

We can nevertheless ask how good is the adiabatic vacuum approximation at sub-Hubble scales as we 
approach this regime. In order to investigate this issue, let us consider a massless test field 
$\phi$ propagating over a \emph{homogeneous} spacetime with metric $g_{\mu\nu}={\rm 
diag}(-1,S^2\gamma_{ij})$. The equation of motion of the field is
\be
\ddot\phi + 3H\dot\phi -S^{-2}\nabla^2\phi = 0\,,
\ee
where $\nabla^2\equiv \gamma^{ij}\partial_i\partial_j$. Note that this equation holds for both  
BI and FL spacetimes -- the main difference being that, in the former, 
$\gamma^{ij}=\gamma^{ij}(\tau)$.

After removing the friction term with a field redefinition $v \equiv  S^{-3/2}\varphi$, the Fourier 
transformed equation of motion becomes 
\be
\label{eqv}
\ddot v_{\mathbf{k}} + \omega^2(k,\tau)v_{\mathbf{k}} = 0\,,
\ee
where we have defined
\be
\label{omegav}
\omega^2(k,\tau) \equiv \frac{k^2}{S^2}-\frac{3}{2}\dot{H}-\frac{9}{4}H^2\,.
\ee

Equation~\eqref{eqv} clearly depends on the directions $k_i$ through Eq.~\eqref{ksquare}, and 
analytical solutions might turn out to be very complicated. But since Eq.~\eqref{eqv} behaves 
formally as a time-independent Schrödinger equation, with the time $\tau$ playing the role of a 
radial distance, we can try to find WKB solutions of the form~\cite{Martin:2002vn}
\be
\label{ansatz}
v_{\mathbf{k}}(\tau) = \frac{1}{\sqrt{2\omega}}\exp\left[{\pm {\rm 
i}\int^\tau\omega\,\dd\tau'}\right]\,.
\ee
This ansatz is a solution of the equation
\be
\ddot v_{\mathbf{k}} + \left(\omega^2-Q_{\rm WKB}\right)v_{\mathbf{k}} = 0\,,
\ee
where
\be
Q_{\rm WKB} \equiv \frac{3}{4}\frac{\dot{\omega}^2}{\omega^2} - 
\frac{1}{2}\frac{\ddot{\omega}}{\omega}\,.
\ee
Obviously,~\eqref{ansatz} will be an approximate solution to \eqref{eqv} as long as $|Q_{\rm 
WKB}|/\omega^2\ll1$. To see if this conditions is met, we can approximate $H$ by 
Eq.~\eqref{hlambda} since, as we have seen, this is a good description of the dynamics when the 
shear still dominates. As $\tau\rightarrow0$, $\dot{H}$ and $H^2$ dominate over the term $k^2/S^2$, 
and one can check that $\omega\sim\tau^{-1}$. We thus deduce that \mbox{$|Q_{\rm WKB}|/\omega^2\sim1$} and 
the WKB approximation fails, regardless of $k$. This behaviour is shown in Figure~\ref{wkbfig} for 
a particular value of the parameter $\alpha$
\begin{figure}[ht]
\begin{centering}
\includegraphics[scale=0.54]{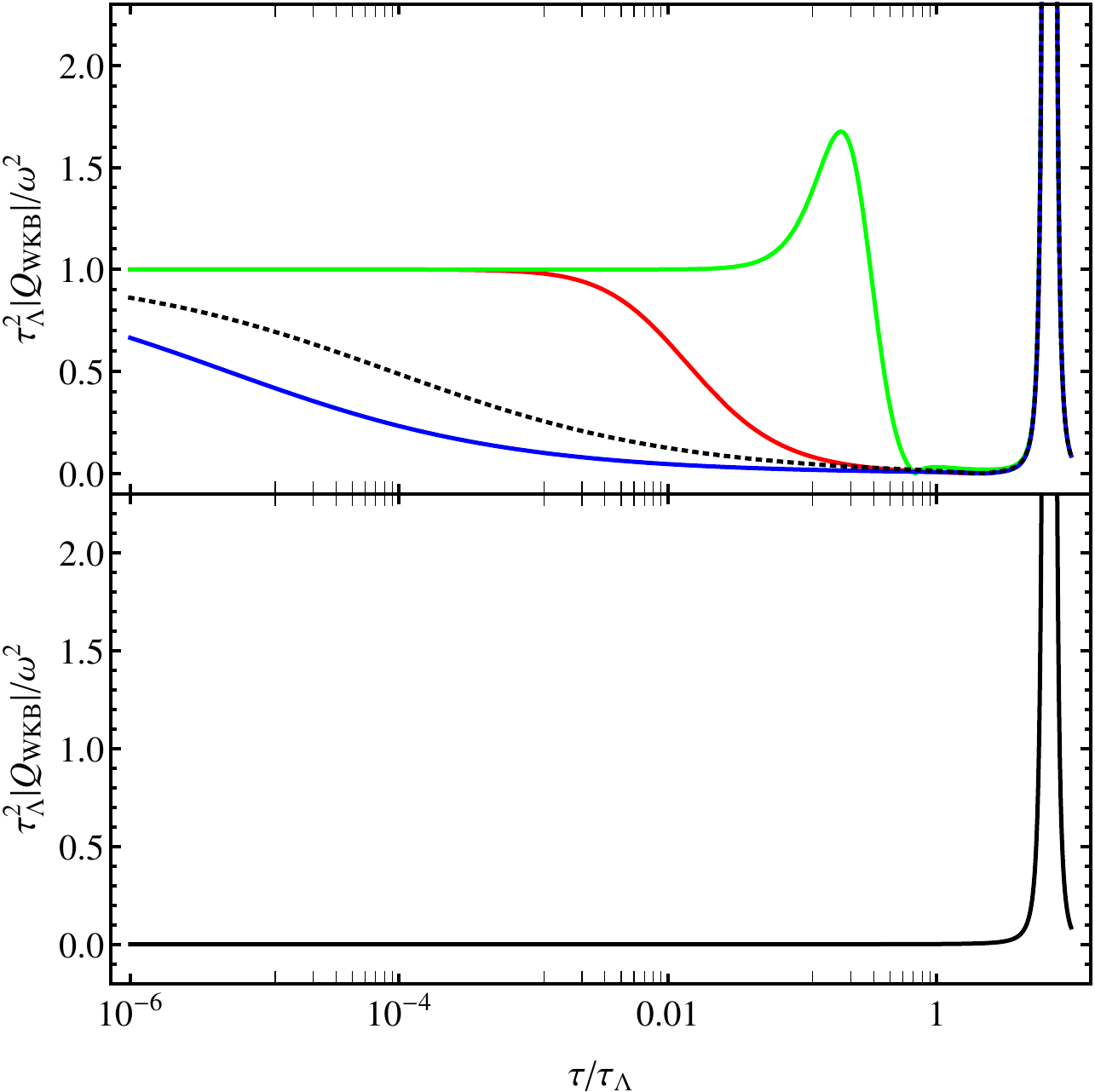}
\par\end{centering}
\protect\caption{Top panel: WKB approximation in BI universes. Continuous lines show the quantity 
$|Q_{{\rm WKB}}|/\omega^2$ as a function of time for a mode $k=5\tau^{-1}_\Lambda$ aligned with 
the $x$, $y$ and $z$ axes (red, green and blue, respectively) and for $\alpha=\pi/4$. The dotted 
line shows a mode with the same modulus distributed equally over the three directions. Note that 
the approximation is good after $\tau\geq\tau_\Lambda$, when the shear has decayed. The bottom 
panel shows the same quantity for a FL universe, and using the same Fourier mode. Here, the 
approximation works throughout the entire past, and the adiabatic vacuum prescription fixes the 
amplitude unambiguously.}\label{wkbfig}
\end{figure}

\subsection{Perturbations generalities}\label{perturb}

The last section focused on the evolution of a test field in an unperturbed background. The 
inclusion of linear matter perturbations coupled to the perturbations of the metric leads to 
additional complications~\cite{Pitrou:2008gk}, but also to potentially new observational features. 
During inflation, these perturbations are described by three canonical variables, one describing the 
scalar perturbations, $v$, and two representing the polarizations of the gravitational waves, 
$\mu_+$ and $\mu_\times$. These variables can be arranged in a three-dimensional vector 
\[
\mathbf{V} = (v(\mathbf{k},\tau),\mu_+(\mathbf{k},\tau),\mu_\times(\mathbf{k},\tau))\,,
\]
whose dynamics is described, in Fourier space, by the following
action~\cite{Pereira:2007yy}
\[
S = \frac{M_p^2}{2}\int\dd t\dd^3k \left(|V'|^2 - k^2|V|^2 + 
\mathbf{V}^\dagger\cdot\mathbf{M}\cdot\mathbf{V}\right)
\]
where $\mathbf{M}$ is an hermitian and non-diagonal matrix which depends on the time, the Fourier mode 
$\mathbf{k}$ and the components of the shear in a given basis. 
Note that the vector perturbations do 
not appear in the action. However, they can no longer be neglected, since they appear as constraints 
relating scalar and tensor modes, and it is crucial to consider them
in intermediary steps when determining the 
form of the action~\cite{Pereira:2007yy}.

When varied, this action leads to
\be
\mathbf{V}''+k^2 \mathbf{V}=\mathbf{M}\cdot\mathbf{V}\,.
\ee
It is the equation ruling the evolution  of three harmonic equations (one
for the scalar mode and two for the tensor modes) coupled by the matrix 
$\mathbf{M}$. These couplings are most important at large scales, and decay at sufficiently small 
scales, as one might expect from the local isotropy of the spacetime. Their general effect is 
twofold: first, since each gravity wave polarization has its own dynamics, their primordial power 
spectra will also differ, mostly at the horizon scale. Second, in the presence of the matrix 
$\mathbf{M}$ the variables $v$, $\mu_+$ and $\mu_\times$ are no longer statistically 
independent~\footnote{However, we are still defining these 
variables as Gaussians.}, even if they are defined independently at $\tau=0$. In particular, this 
implies that the matter correlation at large scales will share power with the correlation of 
gravitational waves, and vice-versa. Nonetheless, the effect of these cross-correlations is 
expected to be small, and in a first analysis we can focus on their individual (self-correlation) 
power spectra:
\be
\label{pspecgen}
{\cal P}_X(\mathbf{k})  = f_X(k)\left[1+\sum_{\ell,m}r^X_{\ell m}(k)Y_{\ell 
m}(\hat{\mathbf{k}})\right]
\ee
where $X$ stands for either $v$, $\mu_+$ and $\mu_\times$. Here, $f_X(k)$ is the monopole of the 
expansion, and represents the isotropic power spectrum. The coefficients $r^X_{\ell m}(k)$ 
characterize the deviation from isotropy, and for this reason they tend to vanish for $\ell\gg1$ 
and $k\gg SH|_{\tau_\Lambda}$, where $\tau_\Lambda$ is the
isotropization time. Furthermore, there are restrictions on the
multipoles  $r^X_{\ell m}(k)$ as we discuss below, and it can have important consequences for the issue of CMB anomalies.

\subsection{Relation to CMB statistical anomalies}\label{cmbanomalies}
Since the assumed isotropy of the universe is one of the central hypotheses of the standard 
cosmological model, local tests of spatial isotropy are a fundamental problem on its 
own~\cite{uzan2010dark}. Nonetheless, further motivation to conduct these tests come from large 
angle features of CMB, which suggest that new physics could be lurking at the horizon scale. Such 
features are known as \emph{statistical anomalies}, and were initially reported by several 
independent groups using WMAP 
data~\cite{Bennett:2010jb,deOliveiraCosta:2003pu,Schwarz:2004gk,Eriksen:2003db,Land:2005ad}. The robustness 
of these anomalies have gained strength after the release of the Planck 
data~\cite{Ade:2013nlj,Ade:2015hxq}, since several of the initial WMAP anomalies have survived the 
completely different pipeline analysis of the Planck 
collaboration~\cite{Copi:2013cya,Copi:2013jna,Bernui:2014gla,Akrami:2014eta} 
(see~\cite{Copi:2010na} for a comprehensive review). Since their initial report, several models in 
which isotropy is explicitly violated have been offered as possible mechanism for the reported 
anomalies~\cite{Jaffe:2005gu,Campanelli:2006vb,Boehmer:2007ut,Rodrigues:2007ny}. It is thus interesting to 
ask what are the generic consequences of anisotropic models to the spectrum of CMB at large scales. To be 
more specific, let us consider the case of 
homogeneous but anisotropic models, such as those resulting from anisotropic Bianchi metrics.

Let ${\cal O}_\mathbf{k}(\tau)$ be one realization of a (Fourier transformed) Gaussian random 
cosmological observable. In a spatially homogeneous but anisotropic spacetime, its ensemble average 
at a fixed time obeys
\be
\label{pspec}
\langle{\cal O}_\mathbf{k}\overline{{\cal O}}_\mathbf{q}\rangle = 
P(\mathbf{k})\delta^{(3)}(\mathbf{k}-\mathbf{q})\,,
\ee
where we have dropped the time dependence for simplicity and the overbar denotes complex 
conjugation. Note that the statistical independence of the scales imposed by the delta 
function is a direct consequence of spatial homogeneity. Spatial anisotropy further demands
the power spectrum $P$ to be a function of the full vector $\mathbf{k}$. From the 
definition~\eqref{pspec} and the properties of the delta function, we also find
\begin{align}
  \langle\overline{{\cal O}_\mathbf{k}\overline{{\cal O}}_{\mathbf{q}}}\rangle & = \nonumber
  \langle\overline{{\cal O}}_\mathbf{k}{\cal O}_\mathbf{q}\rangle\nonumber \\
  & = P(\mathbf{q})\delta^{(3)}(\mathbf{q}-\mathbf{k})\nonumber \\
  & = P(\mathbf{k})\delta^{(3)}(\mathbf{k}-\mathbf{q})\nonumber \\ 
  & = \langle{\cal O}_\mathbf{k}\overline{{\cal O}}_\mathbf{q}\rangle\nonumber\,,
\end{align}
which shows that $P(\mathbf{k})$ is real. It implies the restriction $\bar
r^X_{\ell m}(k)=(-1)^m r^X_{\ell \,-m}(k)$ on Eq.~\eqref{pspecgen}. Furthermore, using the reality 
condition ${\cal O}_\mathbf{k}={\overline{\cal O}}_{-\mathbf{k}}$ together with the last result, we 
find
\begin{align}
\langle{\cal O}_\mathbf{k}\overline{{\cal O}}_\mathbf{q}\rangle & = 
\langle\overline{{\cal O}}_\mathbf{k}{\cal O}_\mathbf{q}\rangle\,\nonumber \\
& = \langle{\cal O}_{-\mathbf{k}}\overline{{\cal O}}_{-\mathbf{q}}\rangle\,\nonumber \\
& = P(-\mathbf{k})\delta^{(3)}(\mathbf{k}-\mathbf{q})\nonumber\,,
\end{align}
which shows that
\be
\label{evenpspec}
P(\mathbf{k})=P(-\mathbf{k})\,.
\ee
Apart from the assumption of null vorticity of the spacetime~\cite{Sundell:2015gra}, this result is 
quite general, and tells us that homogeneous but anisotropic models respect parity.  It implies that 
any $r^X_{\ell m}(k)$ vanishes when with an odd $\ell$. Interestingly, some of the reported CMB 
anomalies look like a genuine violation of 
parity. This is the case, for example, of the quadrupole-octopole alignment~\cite{Land:2005ad}, which 
suggests a temperature covariance matrix of the form $\langle a_{2 m}\overline{a}_{3m}\rangle$. 
However, at large scales, the radiation transfer starting from initial
conditions whose power spectrum respects Eq. \eqref{evenpspec} can only produce 
even-even and odd-odd correlations~\cite{Abramo:2010gk,Pullen:2007tu}. Another example is the 
observed north-south asymmetry in the CMB maps, which is usually modelled with a power spectrum 
of the form $P(\mathbf{k})=P(k)(1+A\hat{\mathbf{k}}\cdot\hat{\mathbf{z}}$), where $A$ is an overall 
amplitude and $\hat{\mathbf{z}}$ is the direction of the north-south 
asymmetry~\cite{Dai:2013kfa,Mazumdar:2013yta}. Clearly, this form violates~\eqref{evenpspec}, and 
thus it cannot result from anisotropy alone.

We thus conclude that, if parity-violating CMB anomalies are indeed a result of new physics, these 
are more likely to result from a break of translation invariance, either 
explicitly~\cite{Carroll:2008br}, or as a consequence of mode-coupling induced by non-gaussian 
statistics~\cite{Schmidt:2012ky,Schmidt:2010gw}.

\section{Alternative models}\label{alternative}

In \S~\ref{SecBackGen}, we have shown that slow-roll inflation classically erases primordial anisotropies, 
so that any initial shear, no matter how large, is quickly diluted by the expansion. However, since 
the details of inflation are still elusive, it is important to point the existence of alternative 
models which could circumvent this fact.

\subsection{Vector inflation}
The lack of primordial ``anisotropic hair'' is a generic feature of Bianchi models in the presence 
of a cosmological constant, and is known as the cosmic no-hair theorem~\cite{Wald:1983ky}. 
However, this theorem can be easily violated if the spacetime is endowed with extra 
anisotropic degrees of freedom. In recent years, several works have addressed the dynamics of 
inflation in the presence of vector 
fields~\cite{Watanabe:2009ct,Yokoyama:2008xw,Ford:1989me,Golovnev:2008cf,Koivisto:2008xf}. The main 
idea behind these models is to preserve some of the primordial anisotropy during inflation, 
so as to potentially produce classical signatures at CMB. 

One of the earliest attempts to inflate the universe with a vector field was conducted in~\cite{Ford:1989me}, 
although the primary motivation was to solve initial condition problems rather than understanding 
classical anisotropies. In this model, a minimally and self-coupled vector theory of the form 
\be
{\cal L} =  -\frac{M^2_p}{4}F_{\mu\nu}F^{\mu\nu} + V\left(A_\lambda A^\lambda\right)
\ee
was used to produce inflation. If the potential $V$ is sufficiently flat, then expansion is almost de Sitter, 
but the final signature is anisotropic due to the presence of a preferred direction in the energy momentum 
tensor. In 2008, Ref.~\cite{Golovnev:2008cf} extended this idea to the case of $N$ randomly 
oriented and non-minimally coupled vector fields, where it was found that inflation can lead to 
reminiscent anisotropies of order $1/\sqrt{N}$. Unfortunately, vector field models of these types 
are plagued with instabilities~\cite{EspositoFarese:2009aj}, and their validity as cosmological 
models is still an ongoing debate.

If vector fields are not the driving source of inflation, they could at least play a role as spectator 
fields. One explicit example arises in the context of supergravity inspired models~\cite{Martin:2007ue}, 
where the expansion is still driven by a canonical scalar field $\varphi$, but this time with a vector field 
coupled to the inflaton by means of a free function $f$
\be
{\cal L}\supset-\frac{1}{4}f(\varphi)^2 F_{\mu\nu}F^{\mu\nu}\,.
\ee
In the context of anisotropic inflation, this idea was explored in~\cite{Watanabe:2009ct}. It was found that, 
for a large class of the coupling function $f$, inflation goes through two slow-roll regimes. Due to the 
presence of the vector field, the ratio\footnote{Their analysis refers to an axi-symmetric BI spacetime, 
which corresponds to $\alpha=\pi/2$ in our conventions. Since in this case two scale factors are equal, and 
the trace-free condition determines the third one, we can neglect the index in $\beta_i$.} $\dot\beta/H$ grows 
during the first slow-roll phase, saturating at values of order a few percent. Indeed, the dynamics of 
the system possesses a tracking solution where the vector field energy density follows that of the inflaton 
during the first slow-roll stage, and remaining nearly constant in the second stage. One can then show 
that, during the second slow-roll phase, the amplitude of the shear scales as
\be
\frac{\sigma}{H}\approx\frac{1}{3}\varepsilon\,,
\ee
where $\varepsilon\equiv-\dot{H}/H^2$ is the slow-roll parameter\footnote{Note that this definition is 
different from ours. See Eq.~\eqref{sr1}.}. Interestingly, this model completely determines the amplitude of 
the shear in terms of the slow-roll parameter. Since the latter is of the order of a few percent, the effect 
of a primordial anisotropy could be bordering on current error bars of CMB temperature spectrum. Moreover, 
since the spacetime is anisotropic, the model also predicts an anisotropic power spectrum in accordance with 
the discussion of last section. More importantly, this model is free from instabilities~\cite{Fleury:2014qfa}, 
offering thus an interesting possibility to test the stages of the universe prior to inflation.

\subsection{Shear-free inflation}

In all the models discussed so far, the anisotropy manifests itself through the spatial 
shear. However, once we are willing to give up rotational invariance, we learn that 
anisotropic expansion is just one possibility. Indeed, we can imagine models where the expansion is 
isotropic (i.e., $\sigma_{ij}=0$), but the curvature of the spatial sections is not. These models are known 
as shear-free cosmologies~\cite{Mimoso:1993ym,Abebe:2015ega}.

The simplest examples of shear-free cosmologies are realized with the Bianchi III (BIII) and 
Kantowski-Sachs (KS) metrics, which have spatial sections of the form $\mathbb{H}^2\times\mathbb{R}$ and 
$\mathbb{S}^2\times\mathbb{R}$, respectively. Evidently, the anisotropy of the spatial sections can only be 
maintained at the cost of an imperfect energy-momentum tensor~\cite{Barrow:1997sy,Barrow:1998ih}. However, 
the shear-free condition strongly constrain its form. Consider for example the evolution equation for the 
shear in 1+3 formalism~\cite{ellis2012relativistic}
\[
\dot{\sigma}_{\mu\nu} + 
\sigma_\mu^{\,\alpha}\sigma_{\alpha\nu}+\frac{2}{3}\theta\sigma_{\mu\nu}-\frac{2}{3}\sigma^2 
h_{\mu\nu}-\frac{1}{2}\pi_{\mu\nu}+ E_{\mu\nu} = 0
\]
where $\theta$ is the expansion scalar, $h_{\mu\nu}$ is the (covariant) spatial metric, $\pi_{\mu\nu}$ the 
matter anisotropic stress and $E_{\mu\nu}$ is the electric part of the Weyl tensor. Clearly, the condition 
$\sigma_{\mu\nu}=0$ leads to $\pi_{\mu\nu}=2E_{\mu\nu}$. The remarkable consequence of 
this choice is that, for the BIII and KS metrics, the background equations are formally the  
Friedmann equations with spatial curvature~\cite{Mimoso:1993ym,Carneiro:2001fz}. Thus, at the 
background level, slow-roll inflation will proceed exactly as in a FL universe, and the spatial 
anisotropy will be diluted as $1/S^2$, as usual. In other words, the observational signatures of 
these models lie entirely in the perturbed sector~\cite{Pereira:2012ma}, and thus in the form of 
the primordial power spectrum.

Shear-free cosmologies have important properties which render them viable cosmological models. 
First, one can check that the choice $\pi_{\mu\nu}=2E_{\mu\nu}$ is a dynamically stable fixed point 
of the background equations. Moreover, during inflation, the equation governing the perturbations 
$\delta\pi_{\mu\nu}$ of the stress tensor has only decaying modes, so that one can make definite 
predictions for the perturbed quantities without worrying with the phenomenological model producing 
such an anisotropic stress~\cite{Pereira:2015pxa}.

The theory of linear perturbations in shear-free cosmologies share similarities with both perturbed 
theories in FL and BI universes. Since there is only one scale factor, perturbative modes do not 
couple dynamically during inflation~\cite{Pereira:2012ma}, and the perturbed equations for the 
scalar degree of freedom is formally identical to the one in FL universes. On the other hand, 
the two tensor degrees of freedom have independent dynamics, so one expect non-trivial signatures 
from gravitational waves. One important difference in shear-free cosmologies result from the 
eigenfunctions of the spatial Laplacian. Due to the presence of spatial curvature, the 
wavelengths of perturbations have an upper limit given by the curvature radius. The existence of 
such limit leads to interesting observational signatures, such as the existence of supercurvature 
perturbations~\cite{Lyth:1995cw}, and their possible effects on CMB through the 
Grishchuk - Zeldovich effect~\cite{GarciaBellido:1995wz}.

\section{Final words}
The fact that inflation is the only viable model for the origin of large-scale structures 
forces us to test its robustness against all sorts of extensions. In this review we have explored 
extensions of slow-roll inflation that accommodate a pre-inflationary anisotropic phase. 
The effect of inflation is to quickly erase classical inhomogeneities, although early 
signatures can survive in the spectrum of primordial quantum fluctuations. Unfortunately, the lack 
of rotational symmetry makes it impossible to define an adiabatic vacuum throughout the 
anisotropic era. Thus, definite predictions can only be trusted for modes sourced at the
onset of inflation, when the shear is nearly zero, but where the WKB approximation is still valid. 
Furthermore, if the number of e-folds exceeds the minimum required to solve cosmological problems, 
no observational signatures from a pre-inflationary anisotropic phase would be left.

On the other hand, it should be noted that the above description corresponds to a frugal 
inflationary model, i.e., single field inflation plus general relativity. In fact, the 
prediction of anisotropic hair can be achieved if one invokes extra degrees of freedom. Although 
pure vector field models are generally plagued with instabilities, one can devise stable models in 
which the vector field is coupled to the inflaton, such as happens in supergravity inspired models. 
In this case, and for a large class of coupling functions, the primordial shear survives 
slow-roll inflation, converging to a final value of the order of slow-roll parameters, and thus 
potentially detectable in CMB maps. 

Another possibility of testing pre-inflationary physics is offered by shear-free cosmological 
models, where the expansion of the universe is isotropic, but spatial curvature is direction 
dependent. Such models represent an explicit demonstration that the observed symmetry of CMB does 
not imply an equally symmetric background, and remind us of the possible perils with standard
symmetry assumptions. Observationally, shear-free models can be tested by the detection 
of supercurvature perturbations, as well as a non-trivial dynamics of gravitational waves.

Finally, we have demonstrated that anisotropic cosmological models without vorticity respect 
parity, as long as spatial homogeneity persists. Thus, since most of the CMB anomalies point 
to a break of parity -- and assuming that they are indeed physical -- they cannot result from a 
break of rotational symmetry alone. This suggest that inhomogeneous cosmological models are more 
likely to explain the existing anomalies.

\begin{acknowledgments}
We thank Jean-Philippe Uzan for inviting us to write this review.
\end{acknowledgments}


\bibliographystyle{h-physrev4}
\addcontentsline{toc}{section}{\refname}\bibliography{review}


\end{document}